\documentclass[prd,twocolumn,amsmath,amssymb,nofootinbib,preprintnumbers,balancelastpage]{revtex4}
\usepackage{graphicx}
\usepackage{hyperref}
\hypersetup{
    pdfnewwindow=true,      
    colorlinks=true,       
    linkcolor=black,          
    citecolor=blue,        
    filecolor=blue,      
    urlcolor=blue           
}
\begin{document}
\title{\Large {\bf{Gauge Theory for Baryon and Lepton Numbers with Leptoquarks}}}
\author{Michael Duerr$^{1}$, Pavel Fileviez P\'erez$^{1}$, Mark B. Wise$^{2}$}
\affiliation{ \\ $^{1}$Particle and Astro-Particle Physics Division \\
Max Planck Institute for Nuclear Physics (MPIK) \\
Saupfercheckweg 1, 69117 Heidelberg, Germany \\ 
$^{2}$California Institute of Technology, Pasadena, CA 91125, USA}
\preprint{CALT 68-2925}
\begin{abstract}
Models where the baryon ($B$) and lepton ($L$) numbers are local gauge symmetries that are spontaneously broken 
at a low scale are revisited. We find new extensions of the Standard Model which predict the existence of fermions that carry both baryon and lepton numbers (i.e.,  leptoquarks). 
The local baryonic and leptonic symmetries can be broken at a scale close to the electroweak scale and we 
do not need to postulate the existence of a large desert to satisfy the experimental constraints 
on baryon number violating processes like proton decay. 
\end{abstract}
\maketitle
{\bf {I. Introduction.}}
In the Standard Model (SM) the baryon and lepton numbers are automatic global symmetries of the renormalizable couplings. 
Non-perturbative quantum effects associated with anomalies break these symmetries but conserve $B-L$.  In order to explain the 
matter--antimatter asymmetry in the Universe, $B-L$ should be broken if we use the standard scenarios for baryogenesis. 

We know that the neutrinos are massive and the lepton number associated to each family of 
leptons in the SM is not conserved. However, it is possible that total lepton number is 
a good symmetry in nature. One can add higher-dimensional operators to the SM, e.g.,
$QQQL/\Lambda_B^2$ and $LLHH/\Lambda_L$,  which have their origin in new degrees of freedom that arise at a high scale where 
the physics is described by a more fundamental theory such as a grand unified theory (GUT). The first of the two operators 
gives proton decay conserving $B-L$, and the second one is responsible for Majorana neutrino masses. 
Unfortunately, in order to satisfy the bounds from proton decay experiments (e.g., $\tau_p > 10^{32-34}$ years\footnote{For a review on proton decay in several scenarios for physics beyond the SM, see Ref.~\cite{review}.}) the relevant scale has to be very high, $\Lambda_B > 10^{15}$ GeV. Hence one needs to postulate 
the existence of a large desert between the weak scale and the scale $\Lambda_B$ where one can understand the origin of these interactions.

In the classical approach based on GUTs, one can compute the 
operators mediating proton decay. Furthermore,  making use of the running of the gauge couplings one  
understands at which scale the larger gauge group is spontaneously broken to the standard model, and hence why the scale $\Lambda_B$ is so large. GUTs make a large number of interesting predictions,  but since they unify quarks and leptons into the same multiplets, baryon and lepton number cannot be defined as  independent symmetries. 

Recently, the authors of Ref.~\cite{paper1} have investigated a different approach in which the baryon and the total lepton 
numbers are independent local gauge symmetries that can be broken at the low scale. Despite the spontaneous breaking of these symmetries the charges of the fields are such that baryon number violating processes are very suppressed  even in the presence of non renormalizable interactions.  Such models provide a way to understand the suppression of baryon and lepton number violating interactions without the necessity of a large desert. Several authors have studied the possibility to gauge $B$ and $L$ as independent symmetries. 
See Refs.~\cite{paper1,paper2,paper3,paper4,Dong} for details (See also Refs.~\cite{old1,old2,old3} for early related studies). 
Unfortunately, all the solutions proposed are in disagreement with the recent constraints from the LHC experiments or with 
cosmological data. 

In this article we revisit the possibility of gauging $B$ and $L$  in an  anomaly-free 
theory  and spontaneously breaking these gauge symmetries at a low scale (e.g., TeV scale). We find that using, what we call, leptoquarks one can cancel 
all anomalies and generate masses for all fields in the theory. In the simplest scenario there is a fermionic dark 
matter candidate, whose stability is an automatic consequence of the gauge symmetry. The new 
fermions in the theory do not induce flavor violation and after symmetry breaking one 
generates $\Delta L=\pm 2, \pm 3$  and $\Delta B=\pm 3$ interactions. Therefore, proton (and baryon number violating neutron) decay is forbidden and there is no need to postulate a large desert. 

This paper is organized as follows: In section II we discuss the conditions coming from the cancellation of the baryonic and leptonic anomalies.
In section III we discuss in detail how to cancel the anomalies in models with fermionic leptoquarks. The simplest viable model is discussed 
in section IV, and we summarize our results in section V.

{\bf{II. B and L as Local Gauge Symmetries.}}
In the Standard Model the baryon and lepton numbers are accidental global symmetries of the Lagrangian but they are not free of anomalies. In order to define a consistent theory where baryon and lepton numbers are local gauge symmetries, all relevant anomalies need to be cancelled. Therefore, the SM particle content has to be extended by additional fermions.
In our notation, the SM fermionic fields and their transformation properties under $SU(3)\otimes SU(2) \otimes U(1)_Y \otimes U(1)_B \otimes U(1)_L$ are given by
\begin{eqnarray}
Q_L  & \sim & (3, 2, \frac{1}{6}, \frac{1}{3}, 0), \ 
u_R  \sim  (3, 1, \frac{2}{3}, \frac{1}{3}, 0), \nonumber \\
 d_R &  \sim & (3, 1, -\frac{1}{3}, \frac{1}{3}, 0), \
 \ell_L  \sim  (1,  2, -\frac{1}{2}, 0, 1), \nonumber \\
\nu_R & \sim & (1, 1, 0, 0, 1) , \ 
e_R  \sim  (1, 1, -1, 0,  1) . \nonumber
\end{eqnarray}  
Here, we already have included the right-handed neutrinos as part of the SM fermionic spectrum.
The purely baryonic anomalies we need to understand are
\begin{eqnarray}
&& \mathcal{A}_1\left(SU(3)^2\otimes U(1)_B\right), \  \mathcal{A}_2\left(SU(2)^2\otimes U(1)_B\right), \nonumber \\
&& \mathcal{A}_3\left(U(1)_Y^2\otimes U(1)_B\right), \  \mathcal{A}_4\left(U(1)_Y\otimes U(1)_B^2\right), \nonumber \\ 
&& \mathcal{A}_5\left(U(1)_B \right), \  \mathcal{A}_6\left(U(1)_B^3\right). \nonumber 
\end{eqnarray}
In the SM, the only non-zero values are $\mathcal{A}_2 = - \mathcal{A}_3 = 3/2$.
In a similar way the purely leptonic anomalies are 
\begin{eqnarray}
&& \mathcal{A}_7 \left(SU(3)^2\otimes U(1)_L\right), \  \mathcal{A}_8 \left(SU(2)^2\otimes U(1)_L\right), \nonumber \\
&& \mathcal{A}_9\left(U(1)_Y^2\otimes U(1)_L\right), \  \mathcal{A}_{10}\left(U(1)_Y\otimes U(1)_L^2\right), \nonumber \\ 
&& \mathcal{A}_{11} \left(U(1)_L \right), \  \mathcal{A}_{12} \left(U(1)_L^3\right), \nonumber
\end{eqnarray}
where only two anomalies are non-zero in the SM with right-handed neutrinos, i.e., $\mathcal{A}_8=-\mathcal{A}_9=3/2$.
In general, one also has to think about the cancellation of the mixed anomalies
\begin{eqnarray}
 &&\mathcal{A}_{13} \left(U(1)_B^2\otimes U(1)_L\right),
 \mathcal{A}_{14} \left(U(1)_L^2\otimes U(1)_B\right), \nonumber  \\
 &&\mathcal{A}_{15} \left(U(1)_Y\otimes U(1)_L\otimes U(1)_B\right), \nonumber
\end{eqnarray}
which of course vanish in the SM. Various solutions to the equations which define the cancellation of anomalies were studied 
in Refs.~\cite{paper1,paper2,paper3,paper4,Dong}:
\begin{itemize}

\item Sequential Family: In Refs.~\cite{paper1,paper2}, a sequential family was proposed, 
where the new quarks have baryon number $-1$ and the new leptons have lepton number $-3$. Unfortunately, this solution is ruled out today because the new quarks get mass from the SM Higgs and change the gluon fusion Higgs production 
by a factor of 9~\cite{Kribs}. This is in disagreement with the recent LHC results. On top of that, 
 the LHC bounds on the masses of the new quarks are strong, and one has Landau poles for the new 
Yukawa couplings in the TeV region.  

\item Mirror Family: In Refs.~\cite{paper1,paper2} the possibility to use mirror fermions was considered, too.
It suffers from the same problems as a sequential family and is also ruled out. 

\item Vector-Like Fermions: To avoid Landau poles close to the electroweak scale, Ref.~\cite{paper4} cancelled  anomalies using vector-like fermions. In this case, anomaly cancellation requires 
that the difference between the baryon numbers of the new quarks is equal to $-1$, while the 
difference between the lepton numbers is $-3$. See Ref.~\cite{paper4} for more details. 

In this setup, the neutrino masses are generated through the Type I seesaw and the new charged leptons get 
mass only from the SM Higgs vacuum expectation value (VEV). Therefore, in this model the lepton number is broken by two units 
and one does not have proton decay. Unfortunately, the new charged leptons change dramatically 
the Higgs branching ratio into gamma gamma~\cite{Mark-Koji}, reducing it by about a factor of 3.
This model disagrees with the recent LHC results where the newly discovered boson is SM-like.         

One can modify this model adding a new Higgs boson with lepton number and generate vector-like masses 
for charged leptons, but one will generate dimension nine operators mediating proton decay, e.g.,
\begin{equation}
{\cal{O}}_9 = c_9 \left( u_R u_R d_R e_R \right) S_B S_L^\dagger S_L^{'}/\Lambda^5.
\end{equation}
Here $S_B \sim (1,1,0,-1,0)$, $S_L \sim (1,1,0,0,-2)$ and $S_L^{'} \sim (1,1,0,0,-3)$. 
Now, assuming that $c_9 \sim 1$, the VEVs of the $S_B, S_L$, and $S_L^{'}$ are around TeV, one finds 
that $\Lambda \geq 10^{7-8}$ GeV. This means that we still have to postulate half of the desert (using a logarithmic scale) in order to 
satisfy the proton decay bounds. Of course, we could also assume that $c_9$ is very small. 

\item Leptoquarks: It is natural to think about cancelling the $B$ and $L$ 
anomalies adding fermionic leptoquarks. This approach was used in Ref.~\cite{Dong} 
where the authors introduced the fields $F_L \sim ({3},{2},0,-1,-1)$, 
$ j_R \sim ({3},{1},\frac{1}{2},-1,-1)$, and $k_R \sim ({3},{1},-\frac{1}{2},-1,-1)$. 
Unfortunately, this model is ruled out by cosmology because one predicts the existence 
of stable charged fields. In the next section, we will elaborate on different possibilities where one 
can avoid this problem.
\end{itemize}

{\bf{III. Fermionic Leptoquarks.}}
As mentioned above, there are different ways to cancel all relevant anomalies to gauge $B$ and $L$.  However, it is difficult to write a consistent  model which is in agreement with collider data and cosmology without postulating the existence of a large desert.
In order to find viable scenarios, we will stick to the particle content listed in Table I, where we use fermionic fields that are singlets or in the fundamental of $SU(2)$. We consider different possibilities for the quantum numbers  of the new fields
under $SU(3)$.
\squeezetable
\renewcommand{\arraystretch}{2}
\begin{table}
 \caption{The extra particle content of the model.}
  \begin{tabular}{cccccc}
\hline
~~Field ~~  &~~ $SU(3)$ ~~ &~~ $SU(2)$~~ &~~ $U(1)_Y$~~ &~~ $U(1)_B$~~ &~~ $U(1)_L$
~~ \\
\hline \hline
 $\Psi_L$   & {\bf N}  & {\bf 2} & $Y_1$ & $B_1=-\frac{3}{2N}$ &
$L_1=-\frac{3}{2N}$ \\
 $\Psi_R$   & {\bf N}  & {\bf 2} & $Y_1$ & $B_2=+\frac{3}{2N}$ &
$L_2=+\frac{3}{2N}$ \\
 $\eta_R$   & {\bf N}  & {\bf 1} & $Y_2$           & $B_3=-\frac{3}{2N}$ &
$L_3=-\frac{3}{2N}$ \\
 $\eta_L$   & {\bf N}  & {\bf 1} & $Y_2$           & $B_4=+\frac{3}{2N}$ &
$L_4=+\frac{3}{2N}$ \\
 $\chi_R$   & {\bf N}  & {\bf 1} & $Y_3$            & $B_5=-\frac{3}{2N}$ &
$L_5=-\frac{3}{2N}$ \\
 $\chi_L$   & {\bf N}  & {\bf 1} & $Y_3$            & $B_6=+\frac{3}{2N}$ &
$L_6=+\frac{3}{2N}$ \\
\hline
\end{tabular}
\end{table}

The $SU(2)^2\otimes U(1)_B$ anomaly can only be cancelled by a field charged under $SU(2)$, 
most conveniently by a doublet. We therefore fix the $SU(2)$ quantum numbers of the new particles to be similar 
to a SM family (one $SU(2)$ doublet and two singlets). To not spoil the SM anomaly cancellation we choose the new fields 
to be vector-like under the SM gauge group. 

Considering the cancellation of the $SU(2)^2\otimes U(1)_B$ anomaly, one finds the condition
\begin{equation}
 B_1 - B_2 = -\frac{3}{N},
\end{equation}
and for simplicity we use $B_1 = -B_2$. The same applies to the corresponding leptonic anomaly, 
and we have
\begin{equation}
  L_1 = -L_2 =  -\frac{3}{2N}\, .
\end{equation}
To cancel the $SU(3)^2\otimes U(1)_B$ anomaly when $N \neq 1$, one needs to impose 
the condition
\begin{equation}
 2(B_1 - B_2)-(B_3 - B_4) - (B_5- B_6) = 0.
\end{equation}
Using
\begin{equation}
 B_4 = - B_3 \text{ and } B_5 = -B_6\, ,
\end{equation}
this reduces to 
\begin{equation}
 2 B_1 - B_3 -B_5 = 0\, , 
\end{equation}
which is most easily cancelled by the choice
\begin{equation}
 B_1 = B_3 = B_5\, .
\end{equation}
Similarly, a good choice is
\begin{equation}
 L_4 = - L_3 \text{, } L_5 = -L_6 \text{, and } L_1 = L_3 = L_5 \, .
\end{equation}
Finally, we have to think about the anomalies with weak hypercharge. With the above used assignment of baryon and lepton numbers, $\mathcal{A}_4$ and $\mathcal{A}_{10}$ 
are always cancelled and do not provide a condition for the hypercharges. From $U(1)_Y^2 \otimes U(1)_B$, we obtain the condition
\begin{equation}
 Y_2^2 + Y_3^2 - 2 Y_1^2 = \frac{1}{2}\, .
\end{equation}
A useful set of solutions for this equation is
\begin{multline}
 (Y_1,Y_2,Y_3) \in \\ \left\{ (\pm \frac{1}{2}, \pm 1,0), (\pm \frac{1}{6}, \pm \frac{2}{3}, \pm \frac{1}{3}), (0,\pm \frac{1}{2},\pm \frac{1}{2} )\right\}\, .
\end{multline}
It is easy to check that---using any of these choices---all baryonic and leptonic anomalies are canceled. 
Since the new particles are vector-like with respect to the SM gauge group, 
the SM anomalies do not pose a problem. Additionally, it can be checked that also the $U(1)_L \otimes U(1)_B^2 $, $U(1)_L^2 \otimes U(1)_B $ and $U(1)_Y\otimes U(1)_L\otimes U(1)_B$
anomalies are cancelled. These could be relevant because we deal with particles charged both under $U(1)_B$ and $U(1)_L$. 

In order to find the scenarios where one avoids a stable electric charged or colored field, we use 
as a guideline that the new fields should have a direct coupling to the SM fermions 
or the lightest particle in the new sector is stable. Now, let us discuss the possible 
scenarios for different values of $N$:
\begin{itemize}
\item $N=1$: If the new fields do not feel the strong interaction, 
the only solution which allows for a stable field in the new sector 
is the one where $Y_1=\pm 1/2$, $Y_2=\pm 1$, and $Y_3=0$. 
Then, if the lightest field is neutral, one can have a dark 
matter candidate. We will discuss this solution in the next section in detail.   

\item $N=3$: If one uses the weak hypercharges  
$Y_1=\pm 1/6$, $Y_2=\pm 2/3$, and $Y_3=\pm 1/3$, a stable colored field can be avoided. Unfortunately, in order to generate vector-like 
masses for the new fields one needs a scalar $S_{BL} \sim (1,1,0,-1,-1)$, and one generates dimension 
seven operators mediating proton decay.   

\item $N=8$: This scenario could be interesting but 
in order to couple the new leptoquarks to the SM fermions we 
need to include extra colored scalar fields. The most attractive way is to add color octet scalars that let the new fermions couple to leptons.  The new colored scalars can decay at one loop to a pair of gluons~\cite{Gresham:2007ri}  after spontaneous symmetry breaking because of couplings in the scalar potential. 
We will not pursue this case further sticking instead to the simplest possible model where $N=1$. 
\end{itemize}

{\bf{IV. Theoretical Framework.}}
Our main goal is to define a consistent anomaly free theory based on the gauge group
$$SU(3) \otimes SU(2) \otimes U(1)_Y \otimes U(1)_B \otimes U(1)_L,$$
 that is consistent with experimental and observational constraints and does not need a large desert to satisfy the proton decay bounds. The simplest of the solutions discussed in the last section is the one with colorless fermions. We discuss it in more detail now. 
The new fermion fields of this model are given in Table I, assuming $N=1$, $Y_1=\pm 1/2$, $Y_2=\pm 1$, and $Y_3=0$ and we focus on this choice of hypercharges in the remainder of the paper. 
We call these fields leptoquarks even though they do not couple 
to quarks and leptons because they have baryon and lepton numbers $\pm 3/2$.
\\
\begin{itemize}

\item Interactions:

Using the quantum numbers of the fields,  the relevant interactions are:
\begin{align}
 - \mathcal{L} & \supset  h_1 \overline{\Psi}_L H \eta_R + h_2 \overline{\Psi}_L \tilde{H} \chi_R  
                        + h_3 \overline{\Psi}_R H \eta_L + h_4 \overline{\Psi}_R \tilde{H} \chi_L  \nonumber \\
                           &+ \lambda_1 \overline{\Psi}_L \Psi_R S_{BL} + \lambda_2 \overline{\eta}_R \eta_L S_{BL} 
                           + \lambda_3 \overline{\chi}_R \chi_L S_{BL}
\nonumber \\
                           &+ a_1 \chi_L \chi_L S_{BL} + a_2 \chi_R \chi_R
S_{BL}^\dagger\, + \text{h.c.}
\end{align}
with $S_{BL} \sim (1, 1,0,- 3,- 3)$. Notice that all interactions proportional to the 
$\lambda_i$ couplings generate vector-like mass terms for the new fermions, while the terms proportional 
to $a_i$ give us the Majorana masses for the neutral fields.
\item Majorana Neutrino Masses:

It is very easy to realize the Type I seesaw~\cite{TypeI} mechanism (even at the weak scale) for neutrino masses including a new Higgs 
$S_L \sim (1, 1,0,0,-2)$, and as usual we have the interactions
\begin{equation}
 -\mathcal{L}_\nu = Y_\nu \overline{\ell}_L \tilde{H} \nu_R + \frac{\lambda_R}{2} \nu_R \nu_R S_L + \text{h.c.}
\end{equation}
\item Symmetry Breaking:

The local baryonic and leptonic symmetries, $U(1)_L$ and $U(1)_B$, are broken by the VEV of $S_{BL}$, while the VEV of $S_L$ only contributes to the breaking of $U(1)_L$.
The fields $S_L$ and $S_{BL}$ can be written as 
\begin{eqnarray}
S_L &=& \frac{1}{\sqrt{2}} (v_L + h_L) + \frac{i}{\sqrt{2}} A_L, \\
S_{BL} &=& \frac{1}{\sqrt{2}} (v_{BL} + h_{BL}) + \frac{i}{\sqrt{2}} A_{BL}. 
\end{eqnarray}
After symmetry breaking the two new physical  scalars $h_L$ and $h_{BL}$ mix with each other and with  the standard model Higgs boson.
\item Fermionic Sector:

After symmetry breaking, in the new sector we have four neutral and four charged chiral fermions, 
$\Psi^0_a$ and $\Psi^{\pm}_b$. It is important to remember that, since the new fermions have baryon number, they don't couple to the SM fermions and one never generates new sources of flavor violation in the SM quark and lepton sectors.

The lightest fermionic field in the new sector is automatically stable and a candidate for the cold 
dark matter of the Universe. Notice that the dark matter stability is a consequence 
of the gauge symmetry and we do not need to impose any discrete symmetry by hand. 
It is important to mention that after the breaking of the local $U(1)_L$ and $U(1)_B$ symmetries 
we get an ${\cal{Z}}_2$ symmetry as a remnant, which is $-1$ for all new fermions and $+1$ for the other fields.

The careful study of the properties of this dark matter candidate is beyond the scope of this article, 
but we would like to mention how one can satisfy the direct detection constraints and achieve 
the right relic density. The dark matter candidate, $\Psi_{LF}$, couples to the 
new neutral gauge bosons in the theory, $Z_1^{'}$ and $Z_2^{'}$, and to the new scalars, 
$h_L$ and $h_{BL}$, one can have the right annihilation cross section if we are close to one of these resonances. 
The direct detection in this case is also through the couplings to the $Z$ and $Z_i^{'}$. Since we have enough 
freedom it is possible to satisfy the direct detection constraints coming from experiments. See Ref.~\cite{Wagner} 
for a recent discussion of the dark matter candidate in models with vector-like leptons. For the impact of these new 
fields on the SM Higgs decays and the constraints from electroweak precision observables see Ref.~\cite{Wagner}. 

\item Constraints from $B$ and $L$ Violating Processes:

Since the new Higgs $S_{BL}$ breaks baryon number in three units, one never generates proton decay.
The field $S_L$ breaks lepton number in two units, so one generates $\Delta L=2$ Majorana mass terms and we 
have the usual constraints coming from neutrinoless double beta decay. 
The lowest-dimensional $B$ and $L$ violating operator, that after symmetry breaking contains just SM fermions, has dimension nineteen
\begin{equation}
{\cal{O}}_{19} = \frac{c_{15}}{\Lambda^{15}} \left(  u_R u_R d_R e_R \right)^3 S_{BL}.
\end{equation} 
Therefore, $B$ and $L$ violating processes are strongly suppressed even if the cut-off of the theory is quite low.
\end{itemize}
%
{\bf{V. Summary.}}
We have proposed a simple extension of the Standard Model where $B$ and $L$ are gauge symmetries  broken at a low scale  and  non renormalizable operators that cause proton decay do not occur. Therefore, there is no need to assume a large desert between the electroweak scale and a scale where additional new physics occurs. Additionally, the new fields needed for anomaly cancellation do not induce new sources of 
flavor violation and one can have a fermionic candidate for the cold dark matter of the Universe.

A potential difficulty for these models is the generation of a cosmological baryon excess~\cite{paper1,paper2}. However it may be possible by making use of accidental global symmetries of the renormalizable couplings in the model or in other ways to generate a non zero baryon asymmetry even though $B$ and $L$ are broken at the low scale. 

 The scenario studied in this article can also be used to 
understand the absence of large  baryon number violating effects in supersymmetric models, where 
typically one uses the symmetry $B-L$~\cite{PRL} as a  framework to understand this issue at a renormalizable level. 

{\textit{Acknowledgments}}:
{\small{M. D.\ is supported by the IMPRS-PTFS. The work of M. B. W. was partly supported by the Gordon and Betty Moore Foundation through Grant \#776 to the Caltech Moore Center for Theoretical Cosmology and Physics and by the  U.S. Department of Energy under contract No. DE-FG02-
92ER40701. }



\begin{thebibliography}{99}

\bibitem{review}
  P.~Nath, P.~Fileviez Perez,
  Phys.\ Rept.\  {\bf 441} (2007) 191.


\bibitem{paper1}
  P.~Fileviez Perez, M.~B.~Wise,
  Phys.\ Rev.\ D {\bf 82} (2010) 011901
   [Erratum-ibid.\ D {\bf 82} (2010) 079901].
  
\bibitem{paper2}
  T.~R.~Dulaney, P.~Fileviez Perez, M.~B.~Wise,
  Phys.\ Rev.\ D {\bf 83} (2011) 023520.
  
\bibitem{paper3}
  P.~Fileviez Perez, M.~B.~Wise,
  Phys.\ Rev.\ D {\bf 84} (2011) 055015.

\bibitem{paper4}
  P.~Fileviez Perez, M.~B.~Wise,
  JHEP {\bf 1108} (2011) 068.

\bibitem{Dong}
  P.~V.~Dong, H.~N.~Long,
  arXiv:1010.3818 [hep-ph].
  
\bibitem{old1}
  A.~Pais,
  Phys.\ Rev.\ D {\bf 8} (1973) 1844;
  S.~Rajpoot,
  Int.\ J.\ Theor.\ Phys.\  {\bf 27} (1988) 689.

\bibitem{old2}
  R.~Foot, G.~C.~Joshi, H.~Lew,
  Phys.\ Rev.\ D {\bf 40} (1989) 2487.

\bibitem{old3}
  C.~D.~Carone, H.~Murayama,
  Phys.\ Rev.\ D {\bf 52} (1995) 484;
  H.~Georgi and S.~L.~Glashow,
  Phys.\ Lett.\ B {\bf 387} (1996) 341.

\bibitem{Kribs}
  G.~D.~Kribs, T.~Plehn, M.~Spannowsky, T.~M.~P.~Tait,
  Phys.\ Rev.\ D {\bf 76} (2007) 075016.


\bibitem{Mark-Koji}
  K.~Ishiwata, M.~B.~Wise,
  Phys.\ Rev.\ D {\bf 84} (2011) 055025.
 
 \bibitem{Gresham:2007ri} 
  M.~I.~Gresham, M.~B.~Wise,
  Phys.\ Rev.\ D {\bf 76}, 075003 (2007).

\bibitem{TypeI}
  P.~Minkowski,
  Phys.\ Lett.\ B {\bf 67} (1977) 421;
  T.~Yanagida,
KEK Report 79-18, Tsukuba (1979);
  M.~Gell-Mann, P.~Ramond and R.~Slansky,
   in {\it Supergravity}, eds.\ P.\ van Nieuwenhuizen et al.,
   (North-Holland, 1979), p.~315;
  S.~L.~Glashow, in {\it Quarks and Leptons}, Carg\`ese, eds.\ M.\ L\'evy et al.,
(Plenum, 1980), p.\ 707;
  R.~N.~Mohapatra and G.~Senjanovi\'c,
  Phys.\ Rev.\ Lett.\  {\bf 44} (1980) 912.
  
\bibitem{Wagner}
  A.~Joglekar, P.~Schwaller, C.~E.~M.~Wagner,
  JHEP {\bf 1212} (2012) 064.

\bibitem{PRL}
V.~Barger, P.~Fileviez Perez, S.~Spinner,
  Phys.\ Rev.\ Lett.\  {\bf 102} (2009) 181802.

   
\end{thebibliography}
\end{document}